\documentclass[journal,twocolumn, english]{IEEEtran}
\usepackage[T1]{fontenc}
\usepackage[latin9]{inputenc}
\usepackage{amsthm}
\usepackage{amsmath}
\usepackage{amssymb}
\usepackage{color}
\usepackage{authblk}
\usepackage{float}
\usepackage{subfig}
\usepackage{graphicx}

\numberwithin{equation}{section}
\numberwithin{figure}{section}
\numberwithin{table}{section}

\theoremstyle{plain}
\newtheorem{thm}{\protect\theoremname}
\theoremstyle{plain}
\newtheorem{prop}[thm]{\protect\propositionname}
\newtheorem{lem}[thm]{\protect\lemmanname}

\newtheorem{cor}[thm]{\protect\corname}

  \providecommand{\conditionname}{Condition}
\providecommand{\definitionname}{Definition}

\makeatother

\usepackage{babel}
\providecommand{\propositionname}{Proposition}
\providecommand{\lemmanname}{Lemma}
\providecommand{\theoremname}{Theorem}
\providecommand{\remarknname}{Remark}
\providecommand{\corname}{Corollary}

\renewcommand\footnotemark{}

\begin{document}

\title{On the Uniqueness of FROG Methods}

\author[1]{Tamir Bendory}
\author[2]{Pavel Sidorenko}
\author[3]{Yonina C. Eldar,~\emph{IEEE Fellow}}

\affil[1]{ The Program in Applied and Computational Mathematics,
Princeton University, Princeton, NJ, USA} \vspace{-2ex}

\affil[2]{Department of Physics and Solid
State Institute, Technion-Israel Institute of Technology, Haifa, Israel} \vspace{-2ex}
\affil[3]{Department of Electrical Engineering, Technion-Israel Institute of Technology, Haifa, Israel}
\maketitle


\begin{abstract} 
The problem of recovering a signal from its power spectrum, called \emph{phase retrieval}, arises in many scientific fields.
One of many examples is ultra-short laser pulse characterization in which the electromagnetic field is oscillating with $\sim10^{15}$ Hz and phase information  cannot be measured directly due to limitations of the electronic sensors. 
Phase retrieval  is  ill-posed  in most cases as there are many
different signals with the same Fourier transform magnitude. To overcome
this fundamental ill-posedness, several measurement techniques are
used in practice. 
One of the most popular methods for complete characterization of ultra-short laser pulses is the 
Frequency-Resolved Optical Gating (FROG). In FROG, the acquired data is the power spectrum of the product of the unknown pulse with its delayed replica. Therefore the measured signal is a quartic function of the unknown pulse.
A generalized version of FROG, where the delayed replica is replaced by a second unknown pulse, is called blind FROG. In this case, the measured signal is quadratic with respect to both pulses. In this letter we introduce and formulate FROG-type techniques. We then show that almost all {band-limited} signals  are determined uniquely, up to trivial ambiguities, by blind FROG measurements (and thus also by FROG), if in addition we  have access to the signals  power spectrum. 
\end{abstract}


\section{Introduction}

{\let\thefootnote\relax\footnotetext{This work was
funded by the European Union's Horizon 2020 research and innovation
program under grant agreement No. 646804-ERCCOG-BNYQ and by the
Israel Science Foundation under Grant no. 335/14. T.B. was partially
funded by the Andrew and Erna Finci Viterbi Fellowship.}}\par

In many measurement systems in physics and engineering one can only acquire the power
spectrum of the underlying signal, namely, its  Fourier transform magnitude.
The problem of recovering a signal from its power spectrum is called \emph{phase retrieval} and it arises in many scientific fields, such as optics, X-ray crystallography, speech recognition, blind channel estimation and astronomy (see for instance, \cite{walther1963question,fienup1982phase,millane1990phase,candes2015phase,shechtman2014phase,jaganathan2015phase} and references therein). 
Phase retrieval for one-dimensional (1D) signals is ill-posed for almost all signals. Two exceptions are minimum phase signals \cite{Huang2015} and sparse signals with structured support \cite{ranieri2013phase,jaganathan2012recovery}. 
Additional information on the sought signal can be used to guarantee uniqueness. For instance, the knowledge of one signal entry  or the magnitude of one entry in the Fourier domain, in addition to the  
the power spectrum, determines almost all signals \cite{beinert2015ambiguities,beinert2016enforcing}.

For general signals,  many algorithms and measurement techniques were suggested to make the problem well-posed. These methods can be classified into two categories. The first utilizes some prior knowledge (if it exists) on the underlying structure of the
signal, such as sparsity (e.g. \cite{ranieri2013phase,shechtman2014gespar,sidorenko2015sparsity}) or knowledge on a portion of the signal (e.g. \cite{fienup1982phase,beinert2015ambiguities,beinert2016enforcing}).
The second  uses techniques that generate
redundancy in the acquired data by taking
additional measurements. These measurements can be obtained for instance using radom masks \cite{candes2014phase,gross2015improved} or by
multiplying the underlying signal with shifted versions of a known reference signal, leading to short-time Fourier measurements \cite{eldar2015sparse,jaganathan2015stft,bendory2016nonconvex}. 

An important application for phase retrieval is ultra-short laser pulse characterization. Since the electromagnetic field is oscillating at $\sim10^{15}$ Hz, phase information  cannot be measured directly due to limitations of the electronic sensors.
To overcome the fundamental ill-posedness of the phase retrieval problem, a popular approach is to use Frequency-Resolved Optical Gating (FROG). This technique measures the power-spectrum of the product of the signal with a shifted version of itself or of another unknown signal. The inverse problem of recovering a signal from its FROG measurements can be thought of as \emph{high-order phase retrieval problem}.
The first goal of this letter is to introduce and formulate  such FROG-type methods.  

Our second contribution is to derive a uniqueness result for FROG-type models. Namely, conditions such that the underlying signal is uniquely determined from the acquired data. A common statement in the optics community, supported by two decades of experimental measurements, is that a laser pulse can be determined uniquely from  FROG measurements if the power spectrum of the unknown signal is also measured. 
To the best of our knowledge, the uniqueness of FROG  methods was analyzed only in \cite{seifert2004nontrivial} under the assumption that we have access  to the full continuous spectrum. In this letter we analyze the discrete setup as it typically appears in applications.

The letter is organized as follows. Section \ref{sec:Mathematical-Model} introduces the FROG problem and 
formulates it mathematically. Section \ref{sec:Main-Results}
presents our uniqueness result, which is proved in Section \ref{sec:proof_thm1}.
Section \ref{sec:Discussion} concludes the letter.

\section{\label{sec:Mathematical-Model} Model and Background}

\begin{figure*}[h]
\centering
{\includegraphics[scale=0.7]{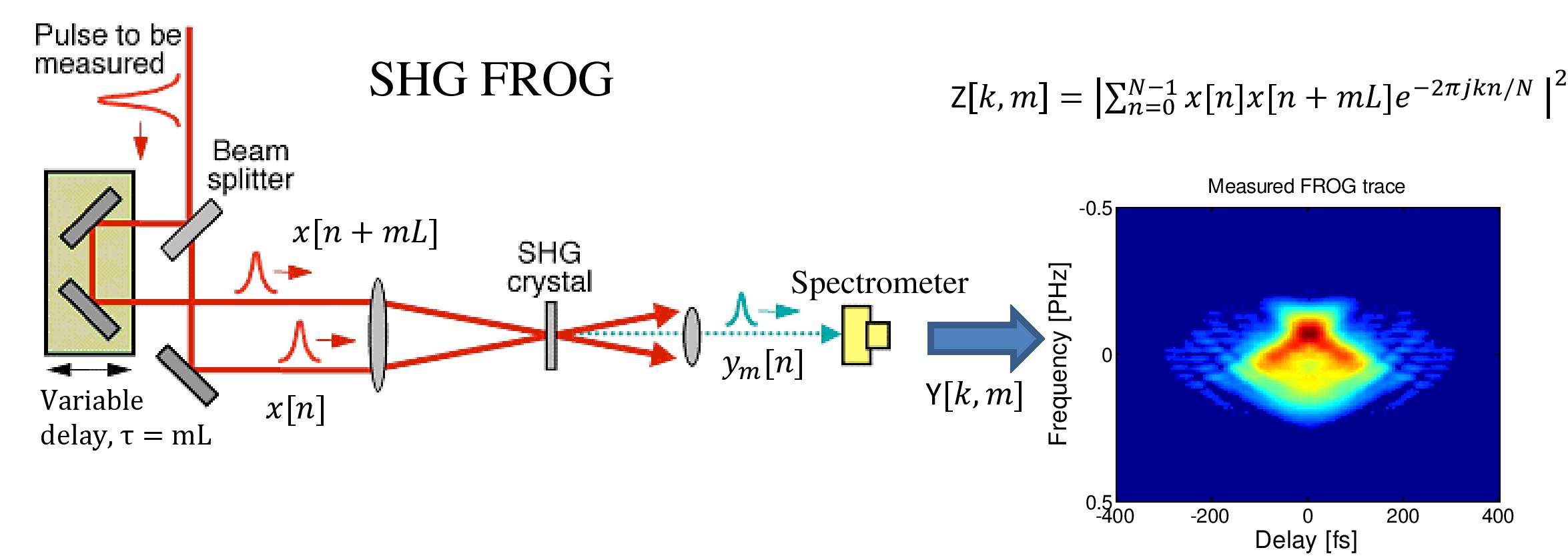}}
\caption{\label{fig:FROG} Illustration of the SHG FROG technique.}
\end{figure*}

We consider two  laser pulse characterization  techniques, called FROG and its generalized version blind FROG. These methods are used to generate redundancy in ultra-short laser pulse measurements. FROG is probably the most commonly used approach for full characterization of ultra-short optical pulses due to its simplicity and good experimental performance \cite{trebino1997measuring,trebino2012frequency}.
A FROG apparatus produces a two-dimensional (2D) intensity diagram of an input pulse by interacting the pulse with  delayed versions of itself in a nonlinear-optical medium, usually using a second harmonic generation (SHG) crystal \cite{delong1994frequency}. This 2D  signal is called a FROG trace and is a quartic function of the unknown signal.  Hereinafter, we consider SHG FROG but other types of nonlinearities exist for FROG measurements. A generalization of FROG in which two different unknown pulses gate each other in a nonlinear medium is called blind FROG. This method can be used to characterize simultaneously two signals \cite{trebino2012frequency,wong2012simultaneously}. In this case, the measured data is referred to as a blind FROG trace and is quadratic in both signals.
 We refer to the problems of recovering a signal from its blind FROG trace and FROG trace as \emph{bivariate phase retrieval} and \emph{quartic phase retrieval}, respectively. Note that quartic phase retrieval is a special case of bivariate phase retrieval where both signals are equal. An illustration of the SHG FROG model is depicted in Figure \ref{fig:FROG}.

In bivariate phase retrieval we acquire, for each delay step $m$, the power spectrum of
\begin{equation} \label{eq:xm}
\mathbf{y}_{m}[n]=\mathbf{x}_1\left[n\right]{\mathbf{x}_2\left[n+mL\right]},
\end{equation}
where $L$ determines the overlap factor between adjacent sections. We assume that $\mathbf{x}_1,\mathbf{x}_2\in\mathbb{C}^N$ are periodic, namely, $\mathbf{x}[i]=\mathbf{x}[N\ell+i]$ for all $\ell\in\mathbb{Z}$. 
The acquired data is given  by
\begin{equation}
\mathbf{Z}\left[k,m\right]=\left|\mathbf{Y}\left[k,m\right]\right|^{2},\label{eq:FROG}
\end{equation}
where
\begin{eqnarray} \label{eq:Y}
\mathbf{Y}\left[k,m\right] & = & \left(\mathbf{F}\mathbf{y}_{m}\right)[k] \nonumber = \sum_{n=0}^{N-1}\mathbf{y}_{m}\left[n\right]e^{-2\pi jkn/N} \nonumber \\
 & = & \sum_{n=0}^{N-1}\mathbf{x}_1\left[n\right]{\mathbf{x}_2\left[n+mL\right]}e^{-2\pi jkn/N},
\end{eqnarray}
and $\mathbf{F}$ is the $N\times N$ DFT matrix.
Quartic phase retrieval is the special case in which $\mathbf{x}_1=\mathbf{x}_2$.

\begin{figure}[h]
	\centering
	\includegraphics[scale=.35]{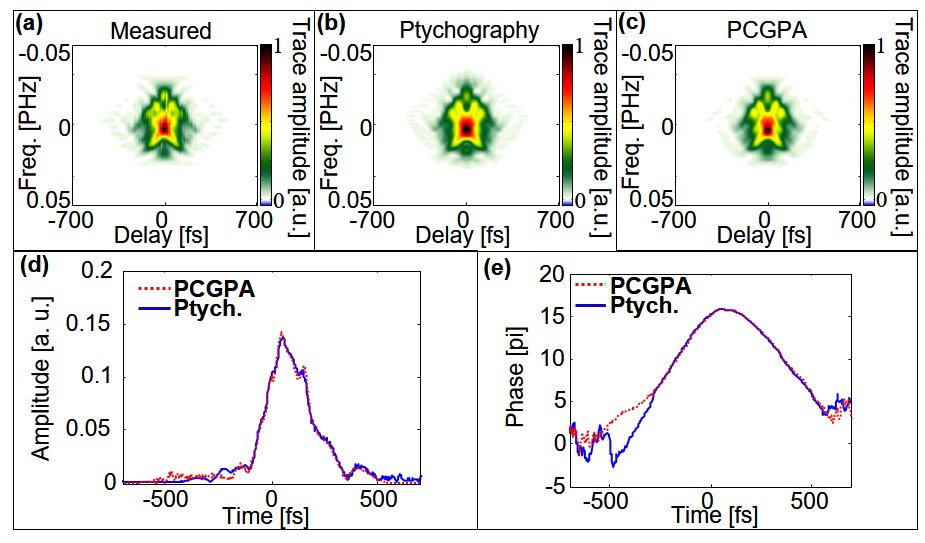}
\protect\caption{\label{fig:Example}  
	Experimental example of  a femtosecond (fs) pulse reconstruction by SHG-FROG. The experiment was conducted with a delay step of 3 fs and 512 delay points. Hence, the complete FROG trace consists of $512\times 512$ data points.
	 The laser pulse was produced by a typical ultrafast Ti-sapphire laser system (1KHz repetition rate, 2 Watt average power).  (a) measured FROG trace (b) recovered trace by alternating projection algorithm for ptychography (Ptych.) proposed in  \cite{sidorenko2016ptychographic} (c) recovered trace by the PCGPA algorithm \cite{kane2008principal} (d) recovered amplitudes by PCGPA and Ptych.  (e) recovered phases by PCGPA and Ptych.}
	
\end{figure}

 Current FROG reconstruction procedures  \cite{fienup1987reconstruction,trebino1993using,millane1996multidimensional} are based on 2D phase retrieval algorithms  \cite{fienup1982phase,fienup2013phase}. One popular iterative  algorithm is the  principal components generalized projections (PCGP) method \cite{kane2008principal}. In each iteration, PCGP performs PCA (principal component analysis, see for instance \cite{jolliffe2002principal}) on a data matrix constructed by a previous estimation. It is common to initialize the algorithm by a Gaussian pulse with random phases.
A recent paper suggests to adopt   ptychographic techniques  where every power spectrum, measured at each delay, is treated separately as a 1D problem \cite{sidorenko2016ptychographic}.  
 In Figure \ref{fig:Example} we present an example for the recovery of a signal from its noisy FROG trace using this algorithm.

In the next section we present our main theoretical results. First, in Proposition \ref{prop:ambiguities} we identify the trivial ambiguities of blind FROG. Trivial ambiguities are the basic operations on the signals $\mathbf{x}_1,\mathbf{x}_2$ that do not change the blind FROG trace $\mathbf{Z}$. Then, we derive a uniqueness result for the mapping between the signals and their blind FROG trace. Particularly, suppose we can measure the power spectra of the unknown signals in addition to the blind FROG trace. 
 We exploit recent advances in the theory of phase retrieval \cite{beinert2016enforcing} and prove that in this case almost all {band-limited} signals are determined uniquely, up to trivial ambiguities. This result holds trivially for FROG as well. 
 The proof is based on the observation that given the signal's power spectrum, the problem can be reduced to standard phase retrieval where both the temporal and spectral magnitudes are known.  

%

\section{\label{sec:Main-Results}Uniqueness Result}

This letter aims at examining under what conditions the measurements $\mathbf{Z}$ determine $\mathbf{x}_1$ and  $\mathbf{x}_2$ 
uniquely. In some cases, there is no way to distinguish between
two pairs of signals, by any method, as they result in the same measurements. The following proposition describes four \emph{trivial ambiguities} of bivariate phase retrieval. The first three  are similar to equivalent results in phase retrieval, see for instance \cite{beinert2015ambiguities}. The proof follows from basic properties of the Fourier transform and is given in the  Appendix.

\begin{prop}
\label{prop:ambiguities}Let $\mathbf{x}_1,\mathbf{x}_2\in\mathbb{C}^N$ and let $\mathbf{y}_m[n]:=\mathbf{x}_1[n]\mathbf{x}_2[n+mL]$ for some fixed $L$. Then, the following signals have the same phaseless bivariate
measurements $\mathbf{Z}[m,k]$ as $\mathbf{x}_1,\mathbf{x}_2$:
\begin{enumerate}
\item multiplication by {global phases} $ \mathbf{x}_1e^{j\psi_1},\mathbf{x}_2e^{j\psi_2}$
for some $\psi_1,\psi_2\in\mathbb{R}$,
\item the {shifted} signal {$$\mathbf{x}_1[n-n_0]\mathbf{x}_2[n-n_0+mL]=\mathbf{{y}}_m[n-n_0]$$
}for some $n_{0}\in\mathbb{Z}$,
\item the {conjugated and reflected} signal { $$\overline{\mathbf{x}_1[-n]}\cdot\overline{\mathbf{x}_2[-n+mL]}=\overline{\mathbf{{{y}}}_m[-n]},$$}
\item modulation, $\mathbf{x}_1[n]e^{-2\pi jk_0n/N}$, $\mathbf{x}_2[n]e^{2\pi jk_0n/N}$ for some $k_0\in\mathbb{Z}$.
\end{enumerate}
\end{prop}

Assume that {one of the signals is band-limited} and that we have access to the power spectrum of the underlying signals $\vert \mathbf{F}\mathbf{x}_1\vert^2$, $\vert \mathbf{F}\mathbf{x}_2\vert^2$ as well as the blind FROG trace $\mathbf{Z}\left[m,k\right]$ of (\ref{eq:FROG}).
In ultra-short pulse characterization experiments {the signals are indeed band-limited \cite{o2004practical}} and the power spectrum of the pulse under investigation is often available, or it can be easily measured by a spectrometer, which is already integrated in any FROG device. 
Inspired by \cite{seifert2004nontrivial}, we show that in this case, the bivariate problem can be
reduced to a standard (monovariate) phase retrieval problem where  both the temporal and the spectral magnitudes are known. Consequently,  we derive the  following result which is proved in the next section.
\begin{thm}
\label{thm:main1}Let $L=1$, and let $\hat{\mathbf{x}}_1:=\mathbf{F}{\mathbf{x}}_1$ and $\hat{\mathbf{x}}_2:=\mathbf{F}{\mathbf{x}}_2$ be the Fourier transforms of  $\mathbf{x}_1$ and  $\mathbf{x}_2$, respectively. Assume that $\hat{\mathbf{x}}_1$ has at least $\left\lceil (N-1)/2\right\rceil$ consecutive zeros (e.g. band-limited signal). 
Then, almost all signals\footnote{By almost all signals, we mean that there may be a set of measure zero for which the theorem does not hold.
} are determined uniquely, up to trivial ambiguities, from the measurements $\mathbf{Z}[m,k]$ and the knowledge of $\vert \hat{\mathbf{x}}_1\vert $ and   $\vert \hat{\mathbf{x}}_2\vert $. By trivial ambiguities we mean that $\mathbf{x}_1$ and $\mathbf{x}_2$ are determined up to global phase, time shift and conjugate reflection. 
\end{thm}
\begin{cor} \label{cor:main}
The same result holds  for quartic phase retrieval in which $\mathbf{x}_1=\mathbf{x}_2$. This model fits the FROG setup.
{\begin{proof}
	The proof follows the proof technique of Theorem \ref{thm:main1} with $\mathbf{x}_1=\mathbf{x}_2$.
\end{proof}}
\end{cor}
%
%
%

\section{Proof of Theorem \ref{thm:main1}} \label{sec:proof_thm1}

The proof is based on the reduction of  bivariate phase retrieval 
to a series of monovariate phase retrieval problems in which both temporal and spectral magnitudes are known \cite{seifert2004nontrivial}. The latter problem is well-posed for almost all signals.

 Let 
 \begin{align*}
 \mathbf{x}_1\left[n\right]&=\frac{1}{{N}}\sum_{\ell=0}^{N-1}\hat{\mathbf{x}}_1\left[\ell\right]e^{2\pi j\ell n/N}, \\
 \mathbf{x}_2\left[n\right]&=\frac{1}{{N}}\sum_{\ell=0}^{N-1}\hat{\mathbf{x}}_2\left[\ell\right]e^{2\pi j\ell n/N},
 \end{align*}
 and \[\mathbf{\delta}[n]:=\begin{cases} 1 &\quad n=0, \\ 0 &\quad \mbox{otherwise}. \end{cases}\]
 Then we have
\begin{eqnarray*}
\mathbf{Y}\left[k,m\right]  &=&  \sum_{n=0}^{N-1}\mathbf{x}_1\left[n\right]\mathbf{x}_2\left[n+m\right]e^{-2\pi jkn/N}\\
 &  =& \frac{1}{N^2}\sum_{n=0}^{N-1}\left(\sum_{\ell_{1}=0}^{N-1}\hat{\mathbf{x}}_1\left[\ell_{1}\right]e^{2\pi j\ell_{1}n/N}\right) \\ && \left(\sum_{\ell_{2}=0}^{N-1}\hat{\mathbf{x}}_2\left[\ell_{2}\right]e^{2\pi jm\ell_{2}/N}e^{2\pi j\ell_{2}n/N}\right)e^{-2\pi jkn/N}\\
& =&   \frac{1}{N^2}\sum_{\ell_{1}=0}^{N-1}\sum_{\ell_{2}=0}^{N-1}\hat{\mathbf{x}}_1\left[\ell_{1}\right]\hat{\mathbf{x}}_2\left[\ell_{2}\right]e^{2\pi jm\ell_{2}/N}\\ && \underbrace{\sum_{n=0}^{N-1}e^{-2\pi j\left(k-\ell_{1}-\ell_{2}\right)n/N}}_{=N\delta\left[k-\ell_{1}-\ell_{2}\right]}\\
  &=& \frac{1}{N}\sum_{\ell=0}^{N-1}\hat{\mathbf{x}}_1\left[k-\ell\right]\hat{\mathbf{x}}_2\left[\ell\right]e^{2\pi jm\ell/N}.
\end{eqnarray*}
Let us denote $\hat{\mathbf{x}}_i\left[\ell\right]=\left|\hat{\mathbf{x}}_i\left[\ell\right]\right|e^{j\boldsymbol{\phi}_i\left[\ell\right]}$ for $i=1,2$, $\mathbf{I}\left[k,\ell\right]=\frac{1}{N}\left|\hat{\mathbf{x}}_1\left[k-\ell\right]\right|\left|\hat{\mathbf{x}}_2\left[\ell\right]\right|$
and $\mathbf{P}\left[k,\ell\right]=\boldsymbol{\phi}_1[k-\ell]+\boldsymbol{\phi}_2[\ell]$. 
Then\footnote{Recall that all indices should be considered as modulo $N$.  Hence,  $\mathbf{Y}\left[k,-m\right]$ is just a reordering of $\mathbf{Y}\left[k,m\right]$.},
\[
\mathbf{Y}\left[k,-m\right]=\sum_{\ell=0}^{N-1}\mathbf{I}\left[k,\ell\right]e^{j\mathbf{P}\left[k,\ell\right]}e^{-2\pi jm\ell/N}.
\]

By assumption, $\vert\hat{\mathbf{x}}_1
\vert$ and $\vert\hat{\mathbf{x}}_2
\vert$ are known and therefore $\mathbf{I}\left[k,\ell\right]$ is known as well. Moreover, note that by assumption, for any fixed $k$, $\mathbf{I}\left[k,\ell\right] $ has at least $ \left\lceil(N-1)/2\right\rceil$ consecutive zeros. Our problem is then reduced to that of recovering  
 the signal $\mathbf{S}\left[k,\ell\right]:=\mathbf{I}\left[k,\ell\right]e^{j\mathbf{P}\left[k,\ell\right]}$ from the knowledge of $\mathbf{Z}[k,-m]$ and $\mathbf{I}\left[k,\ell\right]$. For fixed $k$, this is a standard phase retrieval problem with respect to the second variable where the temporal magnitudes are known.  To proceed,  we  state the finite-discrete version of Theorem 3.4  from  \cite{beinert2016enforcing}:
 
\begin{lem}
\label{thm:quad} Let $t\in\left[0,\dots,N-1\right]\backslash\left\{ (N-1)/2\right\} $ and let $\mathbf{u}\in\mathbb{C}^{N}$ be such that $\mathbf{u}$ has at least $\left\lceil(N-1)/2\right\rceil$ consecutive zeros. Then,  almost every complex signal $\mathbf{u}$  is determined uniquely
 from the magnitude of its Fourier transform and $\left|\mathbf{u}\left[N-1-t\right]\right|$
up to to global phase.  
\end{lem}


 Lemma \ref{thm:quad} implies that $\mathbf{Z}\left[k,-m\right]$
and $\mathbf{I}\left[k,\ell\right]$ determine, for fixed $k$,  almost all $\mathbf{P}\left[k,\ell\right]$ up to global phase.
So, for all $k $, $\mathbf{P}\left[k,\ell\right]$ is determined up to an arbitrary function $\boldsymbol{\psi}[k]$. 
We note that while  Lemma \ref{thm:quad} requires only one sample of $\mathbf{I}\left[k,\ell\right]$ to determine $\mathbf{S}\left[k,\ell\right]$ uniquely,  $\mathbf{I}\left[k,\ell\right]$ does not determine $\vert \hat{\mathbf{x}}_1\vert$ and $\vert \hat{\mathbf{x}}_2\vert$ uniquely. For this reason, we need the full power spectrum of the signals in addition to the blind FROG trace.


Next, we will show that   
\begin{align}  \label{eq:P_tilde}
\mathbf{\tilde{P}}\left[k,\ell\right]&=\mathbf{P}\left[k,\ell\right]+\boldsymbol{\psi}[k] \\
&= \boldsymbol{\phi}_1[k-\ell]+\boldsymbol{\phi}_2[\ell]+\boldsymbol{\psi}[k], \nonumber
\end{align}
 determines $\boldsymbol{\phi}_1, \boldsymbol{\phi}_2$ and $\boldsymbol{\psi}$ up to affine functions. {Note that generally \eqref{eq:P_tilde} may include additional terms of $2\pi s[k,\ell]$ for some integers $s[k,\ell]\in\mathbb{Z}$. However, phase wrapping is physically meaningless since it will not change the light pulse \cite[Section 2]{trebino2012frequency}. 
}  
  
   The relation (\ref{eq:P_tilde}) can be written using matrix notation. Let $\mathbf{\tilde{P}}_{vec}\in\mathbb{R}^{N^2}$ be a column stacked version of $\mathbf{\tilde{P}}$ and let 
\begin{equation*}
\mathbf{v}:=\begin{bmatrix}
\boldsymbol{\phi}_1 \\  \boldsymbol{\phi}_2 \\ \boldsymbol{\psi}
\end{bmatrix} \in \mathbb{R}^{3N}.
\end{equation*}
Then we obtain the over-determined linear system
\begin{equation} \label{eq:P_vec}
\mathbf{\tilde{P}}_{vec} = \mathbf{A}\mathbf{v},
\end{equation}
where $\mathbf{A}\in\mathbb{R}^{N^2\times 3N}$ is the matrix that relates  $\mathbf{v}$ and $\mathbf{\tilde{P}}_{vec}$ according to  (\ref{eq:P_tilde}). 

We aim at identifying the null space of the linear operator $\mathbf{A}$.
 To this end, suppose that there exists another triplet $\boldsymbol{\tilde{\phi}}_1, \boldsymbol{\tilde{\phi}}_2, \boldsymbol{\tilde{\psi}}$ that solves the linear system, i.e.
\begin{align*}  
\mathbf{\tilde{P}}\left[k,\ell\right] = \boldsymbol{\tilde{\phi}}_1[k-\ell]+\boldsymbol{\tilde{\phi}}_2[\ell]+\boldsymbol{\tilde{\psi}}[k], 
\end{align*}
{for all} $k$ and $\ell$. Let us denote the difference functions by $\mathbf{d}_1:= \boldsymbol{\phi}_1 - \boldsymbol{\tilde{\phi}}_1, \mathbf{d}_2:= \boldsymbol{\phi}_2 - \boldsymbol{\tilde{\phi}}_2 $ and $\mathbf{d}_3:= \boldsymbol{\psi} - \boldsymbol{\tilde{\psi}}$.
Then, we can directly conclude that for all $k,\ell$ we have
\begin{equation} \label{eq:d}
\mathbf{d}_1[k-\ell] + \mathbf{d}_2[\ell] + \mathbf{d}_3[k]=0.   
\end{equation}
 Particularly, for $k=0$ and $\ell=0$ we obtain the relations
 \begin{equation} \label{eq:d0}
 \begin{split} 
& \mathbf{d}_1[-\ell] + \mathbf{d}_2[\ell] + \mathbf{d}_3[0]=0, \\   
& \mathbf{d}_1[k] + \mathbf{d}_2[0] + \mathbf{d}_3[k]=0.    
\end{split} 
 \end{equation}
 Plugging (\ref{eq:d0})  into (\ref{eq:d}) (and replace $-\ell$ by $\ell$) we have
\begin{equation*}
\mathbf{d}_1[k+\ell] = \mathbf{d}_1[\ell]+\mathbf{d}_1[k] +\mathbf{d}_3[0] + \mathbf{d}_2[0].   
\end{equation*}
Hence, we conclude that $\mathbf{d}_1$ is an affine function of the form $\mathbf{d}_1[k] = ak-\mathbf{d}_3[0] - \mathbf{d}_2[0]$ for some scalar $a$. { We can also derive that $\mathbf{d}_2[k]=ak+\mathbf{d}_2[0]$ and $\mathbf{d}_3[k]=-ak+\mathbf{d}_3[0]$}. This implies that the null space of $\mathbf{A}$ contains  those affine functions.  
We can compute the phases  by $\mathbf{v}=\mathbf{A}^{\dagger}\mathbf{\tilde{P}}_{vec}$, where 
$\mathbf{A}^{\dagger}$ is the Moore-Penrose pseudoinverse.

To complete the proof, we recall that $\boldsymbol{\phi}_i,\thinspace i=1,2,$  are the phases of the Fourier transforms of $\mathbf{x}_i$. As we can estimate the phases up to affine functions, we can only determine $\mathbf{\hat{x}}_i[k] = \vert \mathbf{\hat{x}}_i[k] \vert e^{j(\boldsymbol{\phi}_i[k]+c_1k+c_2)}$ for some constants $c_1$ and $c_2$. This unknown affine function reflects the global phase and the translation ambiguities. Specifically, the term $e^{jc_1k}$ reflects translation by $c_1$ indices and  the $e^{jc_2}$  product by a global phase. The conjugate-reflectness ambiguity arises from the fact
that both the blind FROG trace and the signals power spectrum
are invariant to this property. This completes the proof.

\section{\label{sec:Discussion}Discussion}

In this paper we analyzed the uniqueness of bivariate and quartic phase
retrieval problems. Particularly, we proposed a uniqueness result showing that given the signals power spectrum, blind FROG trace determines almost all signals up to trivial ambiguities for $L=1$. 
Nevertheless,  it was shown experimentally and numerically \cite{sidorenko2016ptychographic} that stable signal recovery is possible with $L>1$. It is therefore important to   investigate the minimal number of measurements which can guarantee uniqueness for FROG and blind FROG.

It is worth noting different FROG nonlinearities. Two examples are third-harmonic generation FROG and polarization gating FROG. In these techniques, the 
measured signal  is modeled as the power spectrum of 
$\mathbf{y}_m[n] =\mathbf{x}^2[n]\mathbf{x}[n-mL]$  and
$\mathbf{y}_m[n] =\mathbf{x}[n]\vert \mathbf{x}[n-mL]\vert$, respectively \cite{tsang1996frequency,trebino1997measuring}. It is interesting to examine the uniqueness of these high polynomial degree phase retrieval problems in different FROG implementations. Another  important application is the so called Frequency-Resolved Optical Gating for Complete Reconstruction of Attosecond Bursts (FROG CRAB), which is based on the photoionization of atoms by the attosecond field, in the presence of a dressing laser field. In this setup, the signal is modeled as the power spectrum of  $\mathbf{y}_m[n] =\mathbf{x}_1[n]e^{j\mathbf{x}_2[n-mL]}$ \cite{mairesse2005frequency}.

\section*{Acknowledgement}
We would like to thank Kishore Jaganathan for many
insightful discussions,  Oren Cohen for  his advice on  ultra-fast laser pulse measurement methods and Robert  Beinert for helpful discussions about \cite{beinert2016enforcing}.

\appendix
\subsection*{Proof of Proposition \ref{prop:ambiguities}} \label{sec:proof_prof}
The proof is based on basic properties of the DFT matrix. Recall that $\mathbf{y}_m[n]:=\mathbf{x}_1[n]\mathbf{x}_2[n+mL]$.
\begin{enumerate}
\item Let $\psi_1,\psi_2\in\mathbb{R}$ and define $\mathbf{x}_1^\psi:=\mathbf{x}_1e^{j\psi_1}$, $\mathbf{x}_2^\psi:=\mathbf{x}_2e^{j\psi_2}$ and $\mathbf{y}_m^\psi[n]:=\mathbf{x}_1^\psi[n]\mathbf{x}_2^\psi[n+mL]$. Hence,  $\mathbf{y}_m^\psi=\mathbf{y}_me^{j(\psi_1+\psi_2)}$ and it is then clear that  $\mathbf{Z}$ is independent of  $\psi_1,\psi_2$.  
\item Let $n_0\in\mathbb{Z}$ and define $\mathbf{\tilde{y}}_m[n]:=\mathbf{y}_m[n-n_0]$. Then, by standard Fourier properties we get \[\left(\mathbf{F}{\mathbf{\tilde{y}}}_{m}\right)[k]=\left(\mathbf{F}\mathbf{y}_{m}\right)\left[k\right]e^{-2\pi jkn_{0}/N},\]
and consequently $\left|\mathbf{F}{\mathbf{\tilde{y}}}_{m}\right|=\left|\mathbf{F}\mathbf{y}_{m}\right|$. 
\item 
By standard Fourier properties we have $\left|\mathbf{F}{\mathbf{\grave{y}}}_{m}\right|=\left|\mathbf{F}\mathbf{y}_{m}\right|$.
\item Let $k_0\in\mathbb{Z}$ and define $\mathbf{x}_1^{k_0}[n]:=\mathbf{x}_1[n]^{-2\pi j k_0n/N}$, $\mathbf{x}_2^{k_0}[n]:=\mathbf{x}_2[n]^{2\pi j k_0n/N}$ and $\mathbf{y}_m^{k_0}[n]:=\mathbf{x}_1^{k_0}[n]\mathbf{x}_2^{k_0}[n+mL]$. Then, $\mathbf{y}_m^{k_0}[n]=\mathbf{y}_m[n]e^{2\pi j mLk_0/N}$. According to the global phase ambiguity, $\mathbf{Z}$ is independent of $k_0$.
This completes the proof. 
\end{enumerate}

\bibliographystyle{ieeetr}
\bibliography{bib}

\end{document}